\documentclass[a4paper,11pt]{article}
\pdfoutput=1 
\usepackage{graphicx} % including PostScript
\usepackage{jheppub} % for details on the use of the package, please see the JINST-author-manual
\usepackage[utf8]{inputenc}
\usepackage{color}
\usepackage{hyperref}
\usepackage{amsmath,amssymb}
\usepackage{amsfonts}  
\usepackage{mathrsfs}
\usepackage{subcaption}
\usepackage{bbm}
\usepackage{pifont}
\usepackage{enumerate} % enumeration
\usepackage{pdflscape}
\usepackage{hhline}  %for command \hhline
\usepackage{multirow} %
\usepackage{booktabs} %for commands such as \toprule
\usepackage{tabulary}
\usepackage[compat=1.1.0]{tikz-feynman}
\usepackage{nicefrac}
\usepackage{braket}
\usepackage{comment}
\usepackage{xspace}
\usepackage{cancel}
\usepackage{nicematrix}
\usepackage{bbold}
\usepackage{esvect}
\usepackage{arydshln}
\usepackage{tikz}
\usetikzlibrary{positioning}
\usepackage[normalem]{ulem}
\usepackage{xcolor}
\usepackage{enumitem}

\newcommand{\eq}{{\rm eq}}
\newcommand{\SM}{{\rm SM}}
\newcommand{\dd}{{\rm d}}
\newcommand{\GeV}{{\rm GeV}}

\newcommand{\vev}[1]{\langle #1 \rangle}
\definecolor{gg}{RGB}{64, 130, 109}
\definecolor{math2}{rgb}{0.880722, 0.611041, 0.142051}

\arxivnumber{} % if you have one

\title{\boldmath Exploring non-equilibrium effects in sequential freeze-in}

\author{Shiuli Chatterjee}
\author{and Andrzej Hryczuk}
\affiliation{National Centre for Nuclear Research, Pasteura 7, 02-093 Warsaw, Poland}

\emailAdd{shiulic@alum.iisc.ac.in}
\emailAdd{andrzej.hryczuk@ncbj.gov.pl}

\abstract{
Freeze-in of multi-component dark sectors is governed not only by the interaction with the thermal plasma, but also by their internal dynamics. Full thermalisation within the dark sector is not guaranteed, raising the question of impact of departures from local thermal equilibrium onto the evolution and ultimately relic abundance and momentum distribution of dark matter. In this work we explore this question in a minimal two-scalar model, which  can give rise to observable signatures in indirect detection and long-lived particle searches at forward physics experiments. Focusing on the phenomenologically viable regions, we analyse the impact of non-thermal evolution on the dark matter abundance, finding deviations of up to an order of magnitude between the full phase-space treatment and the traditional number-density approach. Our results highlight the importance of phase-space level computation for accurate freeze-in predictions and further motivate dedicated numerical tools for studying the evolution of multi-component dark sectors at the phase space level.
}

\begin{document}
\maketitle
\flushbottom

%%%%%%%%%%%%%%%%%%%%%%%%%%%%%%%%%%%%%%%%%%%%%%%%%%%%%
\section{Introduction}
\label{sec:intro}
%%%%%%%%%%%%%%%%%%%%%%%%%%%%%%%%%%%%%%%%%%%%%%%%%%%%%

The identity of dark matter (DM) remains one of the major open questions in particle physics and cosmology. Indeed, the only property of DM actually measured so far is its abundance. A combination of several datasets allows for its determination, within the $\Lambda$CDM model, to a percent-level precision $\Omega h^2=0.12\pm0.0012$ \cite{Planck:2018vyg}. Nevertheless, not only the nature of DM, but also how the observed abundance was formed, is yet unknown. 

There exist a number of possible mechanisms of DM production that lead to the same total abundance but differ in the evolution in the history of the early Universe. Although not directly measurable, these differences need to be properly understood and accounted for when making predictions for the relic abundance of the DM candidate of a given particle physics model. 
In particular, one consequential aspect of this evolution is whether the DM momentum distribution $f(p)$ traces a thermal form $f_{\rm \eq}(p)$, and, if not, how significant for observable quantities are the resulting departures from equilibrium, especially for the relic abundance and large-scale structure formation.

The expectation that $f(p)\propto f_{\eq}(p)$ throughout all the relevant epochs of the evolution of DM population is a common, yet in many instances oversimplified, assumption. It is well justified in vanilla weakly interacting massive particle (WIMP) freeze-out models without any sharp features in the annihilation cross section (\textit{e.g.}~resonances or strong thresholds), or in a direct freeze-in where DM density always remain orders of magnitude below the one in equilibrium with the Standard Model (SM) plasma.\footnote{
In freeze-in scenarios, the feeble interactions between DM and the SM typically prevent DM from ever reaching kinetic equilibrium with the SM bath. Therefore, DM phase space distribution is expected to be non-thermal. However, the distribution is often found to be \textit{quasi-thermal}, since its shape is nevertheless inherited from thermal bath particles.
} However, beyond these classes of models this approximation may or may not be accurate enough. 

There exist a number of scenarios where it has been shown that departure from equilibrium distributions have a significant phenomenological impact. For freeze-out models: early kinetic decoupling due to kinematical features \cite{Duch:2017nbe,Binder:2017rgn}, heavy scattering partners \cite{Binder:2021bmg} or strong semi-annihilations \cite{Kamada:2017gfc,Cai:2018imb,Hektor:2019ote}, equilibrium disruption in multicomponent dark sectors \cite{Hryczuk:2022gay,Chatterjee:2025vdz}, in dynamical dark matter models \cite{Dienes:2020bmn}, in secluded dark sectors \cite{Beauchesne:2024zsq}, or transition regime between freeze-out and freeze-in \cite{Du:2021jcj}. While for freeze-in additionally: semi-production scenarios \cite{Bringmann:2021tjr,Hryczuk:2021qtz,Bhatia:2023yux}, some low reheating temperature cases \cite{Feiteira:2026qme}, stimulated emission in bosonic DM \cite{Sakurai:2024apm}, modified expansion history \cite{DEramo:2025fvy}, light DM from freeze-in or superWIMP \cite{DEramo:2020gpr,Decant:2021mhj}, sequential freeze-in \cite{Belanger:2020npe}, and even in axion \cite{Badziak:2024qjg} and axino models \cite{Bae:2017dpt}.

Despite these developments, a systematic understanding of non-equilibrium effects in freeze-in scenarios, particularly at the level of phase-space distributions, remains limited.
In this work we perform a detailed case study of a minimal model with two-component dark sector undergoing different realizations of freeze-in production: direct, sequential and/or followed by a dark freeze-out. The model not only acts as a testing ground for how large are the effects of departure from kinetic equilibrium in a freeze-in type scenario, but also is an interesting dark matter realization in its own right with potential for detection in the upcoming experimental searches.

In particular, we focus on scenarios where direct production of DM from the SM is suppressed, and the DM abundance is generated via an intermediate dark sector particle, i.e.~through sequential production~\cite{Hambye:2019dwd}. In this case, the final DM abundance is necessarily sensitive to the phase-space distribution of the intermediate particle~\cite{Belanger:2020npe}. While the presence of non-thermal effects is straightforward to establish, their quantitative impact on the final DM abundance is difficult to estimate \textit{a priori}. Moreover, this impact is expected to be both model- and parameter-space dependent.

In this work we focus on a scalar dark sector with feeble couplings to the SM. The coupling between the two dark sector particles, however, is allowed to vary freely. This setup then admits also the possibility of detection through indirect detection channels (see, \textit{e.g.}~\cite{Siqueira:2019wdg, Heikinheimo:2018duk}). 
Additionally, the existence of another state in the dark sector, decaying to SM particles, can lead to signals in searches for long lived particles at forward physics experiments (see for \textit{e.g.}~\cite{Yin:2025gli}, for a review see \cite{Anchordoqui:2026kpw}).

The paper is organised as follows. In section~\ref{sec:model}, we introduce the model that we consider. Section~\ref{sec:scan} presents its phenomenology including the parameter-space scan and the experimental signatures. In section~\ref{sec:relic}, we outline the different approaches to solving the Boltzmann equation for the DM abundance, together with their underlying assumptions and limitations. We then study the impact of non-thermal evolution in the dark sector on the phenomenological signatures discussed above, using representative benchmark points from the scan, and examine its dependence on the interaction strengths of the model, in section~\ref{sec:BMs}. We conclude in section~\ref{sec:conclusions}.

%%%%%%%%%%%%%%%%%%%%%%%%%%%%%%%%%%%%%%%%%%%%%%%%%%%%%
\section{The model}
\label{sec:model}
%%%%%%%%%%%%%%%%%%%%%%%%%%%%%%%%%%%%%%%%%%%%%%%%%%%%%

The simplest possible model that incorporates two scalar particles in the dark sector, and arguably also the simplest realization of sequential freeze-in, is a model with two real scalars coupled to the visible sector through a Higgs portal. In the most economical version, both scalars are odd under separate $\mathbb{Z}_2$ symmetries, ensuring their stability.
However, such a choice leads to two-component dark matter and, more importantly, in the freeze-in scenario this would result in only feeble interactions with the SM and meagre prospects of any form of detection. 
We therefore assume that one of the $\mathbb{Z}_2$ symmetries is explicitly (softly) broken, yielding a \textit{mediator} state $\phi$ that mixes with the Higgs $h$ and is unstable, decaying into lighter SM states. The second scalar, $S$, is stable and plays the role of DM. 
Such models are very well known in the literature and provide a rich phenomenology for collider searches and astroparticle probes (see \textit{e.g.}\,\cite{Lebedev:2021xey} for a review).

In particular, the scalar potential of the model is,\footnote{This is nearly the most general scalar potential consistent with the adopted symmetries. The missing terms, $B \phi S^2$ and $g_\phi \phi^3$, are disregarded for simplicity: they do not lead to any \textit{new} phenomenologically relevant processes, but complicate the amplitudes for the ones studied in this work. Note that the former could lead to a decay of $\phi\to SS$, but only when $m_\phi>2m_S$, which is a parameter region where present-day $SS \to \phi\phi$ would not be possible thus hindering ID prospects. The latter term, could lead to cannibal and semi-annihilation type reactions, that are however crucially dependent on the value of the coupling being large.
Both terms could be included in the analysis, but at the cost of significant additional complexity without materially affecting the aspects relevant to this work.}
\begin{eqnarray}
   V &=& \mu_h^2 H^\dagger H - \frac{\lambda_h}{2}(H^\dagger H)^2 \nonumber \\ && - A\phi H^\dagger H - \frac{\lambda_{\phi h}}{2} \phi^2 H^\dagger H - \frac{1}{2}\mu_\phi^2 \phi^2 - \frac{\lambda_\phi}{4!}\phi^4  \\ &&- \frac{1}{2}\mu_S^2S^2-\frac{\lambda_S}{4!}S^4 - \frac{\lambda_{S h}}{2} S^2 H^\dagger H - \frac{1}{4}\lambda_{S \phi} S^2\phi^2, \nonumber
\end{eqnarray}
where $H$ is the SM Higgs doublet.
Below $T =T_{\rm EWPT} \sim 150~\GeV$ both $H$ and $\phi$ acquire their VEVs, such that in the unitary gauge $H = (0, v+h_H)/\sqrt{2} $ and $\phi = v_{\phi} + \varphi$. The physical scalars appear as the combinations of $h_H$ and $\varphi$
\begin{equation}
    \begin{pmatrix} h \\ \phi \end{pmatrix} = \begin{pmatrix} \cos{\theta} & -\sin{\theta} \\ \sin{\theta} & \cos{\theta} \end{pmatrix} \begin{pmatrix} h_H \\ \varphi \end{pmatrix} \, ,
\end{equation}
where we now denote the mass eigenstate of the new scalar by $\phi$. Assuming that $\vev{H^\dagger H} = v^2/2$  with $v=246~\GeV$ at the minimum of the scalar potential, and $A\ll v$ and $\lambda_{\phi}\ll 1$, the minimisation conditions yield
\begin{eqnarray}
v_{\phi} &\approx& -\frac{Av^2}{2m^2_{\phi}}, \\
\lambda_h &\approx& 2 \left(\frac{\mu_h}{v}\right)^2 + \frac{A^2}{m^2_{\phi}}
,
\end{eqnarray}
with $m^2_{\phi} \approx \mu^2_{\phi} + \lambda_{h\phi}v^2/2$. We focus on the case where $h$ is essentially the SM Higgs with the mass $m_h \approx 125~\GeV$ and $\phi$ is weakly coupled to it, corresponding to the mixing between the Higgs and the mediator $\theta \ll 1$, which in terms of the model parameters is given by:
\begin{equation}
 \sin\theta = 
 \frac{A v}{
   m_h^2 - m_\phi^2}\left(1 - \frac{\lambda_{h \phi} v^2}{2m_\phi^2}\right)  .
\end{equation}

It is worth mentioning that the initially discarded cubic terms ($\phi^3$ and $\phi S^2$) may be regenerated after $\phi$ gets a VEV, as well as quartic terms 
$h \phi^3$ and $h S^2 \phi$.
However, the couplings in front of these terms are proportional to ``small coupling" $\times ~ v_\phi \sim$ ``small coupling squared", and are thus negligible.\footnote{
This argument might break down in some parameter regions. In such cases it is worth remembering that for the cubic terms, it is always possible to introduce back the originally neglected terms with couplings that cancel the re-generated ones.   
The quartic terms, although not cancellable, contribute only to additional inelastic processes that are suppressed by small couplings and are neglected in the present analysis.} The mass of $S$ also gets a correction, $m_{S} = \sqrt{\mu^2_{\chi} + \lambda_{S\phi} v^2_{\phi}/2}$, which again is negligible in our setup since $\lambda_{S\phi} < 1$ and $v_\phi \ll v$.

For small mixing angles, the mediator $\phi$ is long-lived. 
At low masses, where only its decay to into a lepton pair and photons are kinematically allowed, the mediator can become long-lived enough to be  considered effectively  stable during the freeze-in period. It subsequently decays, with a lifetime determined by $\Gamma_\phi \approx \theta^2 \Gamma_{h\to \mathrm{SM}}(m_\phi)$. 
Such scenarios can, however, be strongly constrained by astrophysical and beam dump limits, as discussed in sec.\,\ref{sec:scan}, and we therefore restrict our discussion to $m_\phi >100~\mathrm{MeV}$.

The introduced setup provides a relatively minimal framework in which sequential freeze-in and its associated non-equilibrium dynamics can be studied consistently. At early times, before EWPT for $T > T_{\rm EWPT}$, $\phi$ and $S$ are predominantly produced via the processes $hh \rightarrow \phi\phi, SS$ with Higgs in the unbroken phase. These are relatively inefficient and, although included in the computations, do not have any significant impact on the evolution. During the EWPT there is a short time window when Higgs bosons can convert to $\phi$ through oscillations~\cite{Heeba:2018wtf}, which however do not give rise to an appreciable contribution in the parameter regions studied here and we do not take them into account. Instead, the main production mode comes in at later times, $T<T_{\rm EWPT}$, when as the system moves to a broken phase Higgs decays ($h \rightarrow \phi\phi,SS$) open up whenever kinematically allowed. The relative branching ratios to $\phi$ and $S$ determine whether the freeze-in production of DM is direct (when $h \rightarrow SS$ is appreciable) or sequential, when the main production process is the conversion $\phi\phi \to SS$.

%%%%%%%%%%%%%%%%%%%%%%%%%%%%%%%%%
\begin{figure}[t]
\centering
\includegraphics[scale=1]{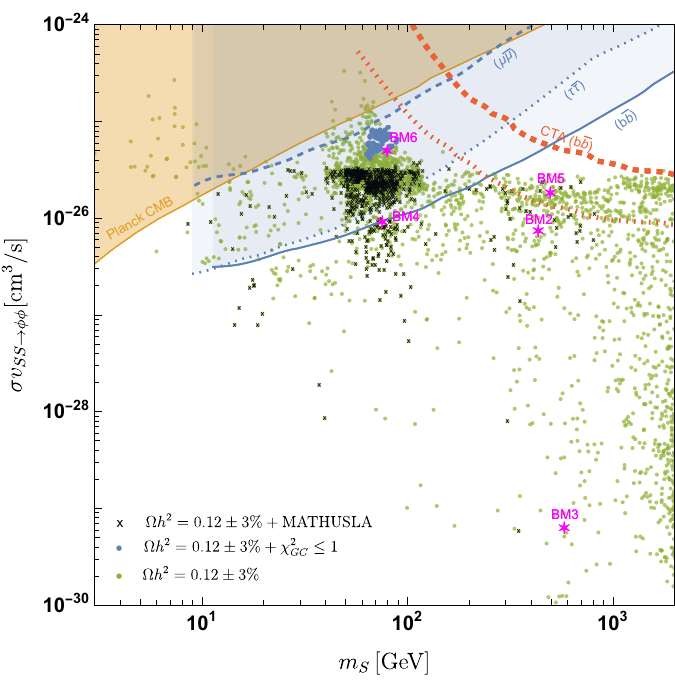}
\caption{Results of the scan showing points that correspond to $\Omega h^2= 0.12\pm 3\sigma=0.12\pm 0.0036$, using the integrated Boltzmann equation for number density evolution (nBE). Superimposed are limits from indirect detection. Existing limits from Planck measurements of the CMB \cite{Planck:2018vyg} (light brown) and Fermi-LAT+MAGIC+HAWC+H.E.S.S.+VERITAS combined analysis of the gamma-ray observations of dSphs \cite{Ahnen:2016qkx} (blue) are shown as shaded regions. The dashed red line indicates the projected sensitivity of CTA for the  $b\bar b$ channel~\cite{Acharyya:2020sbj}, while the dotted line shows an earlier, more optimistic projection~\cite{Wood:2013taa}.
All limits, except those from CMB, are rescaled to account for the cascade annihilation. The magenta points indicate selected benchmark scenarios representative of different production regimes.
}
\label{fig:scanID}
\end{figure}

%%%%%%%%%%%%%%%%%%%%%%%%%%%%%%%%%%%%%%%%%%%%%%%%%%%%%
\section{Phenomenological Scan}
\label{sec:scan}
%%%%%%%%%%%%%%%%%%%%%%%%%%%%%%%%%%%%%%%%%%%%%%%%%%%%

In order to explore the phenomenology of the model, we perform a scan over the six-dimensional parameter space defined by $m_\phi, m_S $, $\lambda_{h\phi},\lambda_{hS}, \lambda_{S\phi}$ and  $\sin \theta$, to find points that satisfy the relic density constraint.
We scan over the parameter space, with  masses and couplings
\begin{gather}
    m_\phi \in [0.1,300] {\ \rm {GeV}}, \quad m_S\in [0.1,3000] {\ \rm {GeV}}, \nonumber\\
    \lambda_{hS}\in [10^{-15},10^{-5}],  \quad\lambda_{h\phi}\in [10^{-15},10^{-5}], \quad\lambda_{S\phi}\in[10^{-4},1],\quad\sin\theta\in[10^{-10},10^{-3}],   \nonumber
\end{gather}
with logarithmic sampling.
The ranges were chosen to cover a wide range of masses for $\phi$ and $S$, with the added condition that $m_\phi<m_h/2$, allowing production of the mediator from Higgs decays, and  $m_S>m_\phi$, choosing one of the two possible mass hierarchies, to allow for prospect of indirect detection via $SS\rightarrow\phi\phi\rightarrow 4\,\SM$. 
While the couplings are chosen with relevance for freeze-in dynamics, the mixing angle is restricted to $\sin\theta<10^{-2}$ to remain consistent with collider constraints \cite{OConnell:2006rsp}. 

\begin{figure}[t]
\centering
\includegraphics[scale=1]{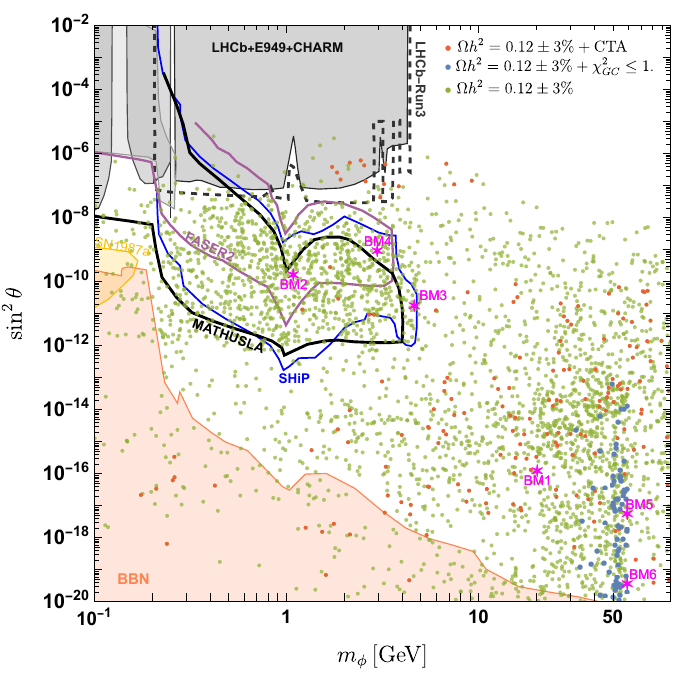}
\caption{Results of the nBE scan showing points with $\Omega h^2= 0.12\pm 3\sigma =0.12\pm 0.0036$  projected onto the $(m_\phi,\sin^2\theta)$ plane.  The superimposed constraints on the mediator $\phi$ are shown as shaded regions: results from CHARM \cite{Bergsma:1985qz}, E949 \cite{Artamonov:2008qb}
and LHCb \cite{Aaij:2016qsm,Aaij:2015tna} recast using \cite{Winkler:2018qyg} (gray); the limits from BBN \cite{Fradette:2017sdd} (orange) and SN1987a \cite{Krnjaic:2015mbs} (yellow). We also show projected sensitivities from future experiments: FASER \cite{Ariga:2018uku} (violet), SHiP \cite{Anelli:2015pba} (blue) and MATHUSLA \cite{MATHUSLA:2022sze} (solid black) and LHCb-Run3 (dashed black) adopted from~\cite{Lanfranchi:2020crw}. 
The magenta points indicate selected benchmark scenarios representative of different production regimes.
}
\label{fig:scanphi}
\end{figure}

The results of the scan are shown in figs.\,\ref{fig:scanID} and \ref{fig:scanphi}, along with phenomenological implications from indirect detection channels and forward physics experiments. All  points shown in the two figures satisfy the observed relic abundance, $\Omega h^2 = 0.12 \pm 3\,\sigma$.
The relic density used in the scan is obtained from the standard number density Boltzmann equation (nBE). The underlying assumption of kinetic equilibrium  in this approximation and its limitations, as well as more general and accurate treatments of the Boltzmann equation are discussed in section \ref{sec:relic}, with the corresponding effects  on the relic abundance analysed in the following sections.

In fig.\,\ref{fig:scanID}, we focus on indirect detection observations. While annihilation into SM particles is typically suppressed in freeze-in models, the present two-scalar setup allows indirect detection through the one-step cascade channel $SS \rightarrow \phi\phi \rightarrow \text{SM}$, with photons produced in the subsequent decays. Since the coupling $\lambda_{S\phi}$ is not required to be feeble by the freeze-in production mechanism, observable indirect-detection signatures can arise. We project all points onto the plane of the  annihilation cross section for $SS\rightarrow\phi\phi$ as function of DM mass, with all other parameters varying. We show constraints from CMB measurements by Planck collaboration~\cite{Planck:2018vyg},  combined limits from the joint analysis of the gamma-ray flux from dwarf spheroidal galaxies (dSphs) performed by Fermi-LAT, MAGIC, HAWC, H.E.S.S.\,and VERITAS~\cite{Fermi-LAT:2025gei}, and projected limits from CTA~\cite{Acharyya:2020sbj,Wood:2013taa}. The latter two limits are reinterpreted for the present case of DM annihilation to SM particles via a one-step cascade, as discussed in section\,\ref{sec:ID}. 

In fig.\,\ref{fig:scanphi}, we focus on the detection prospects of the mediator. For small $\phi$ masses, the decay channels are restricted to leptons and photons, leading to long lifetimes and potential observability at forward physics experiments. Projecting onto the plane of mediator mass and its mixing with Higgs, we show the  current constraints from CHARM \cite{Bergsma:1985qz}, E949 \cite{Artamonov:2008qb} and LHCb \cite{Aaij:2016qsm,Aaij:2015tna} recast using \cite{Winkler:2018qyg}, together with 
projected sensitivities from future experiments FASER\,\cite{Ariga:2018uku}, SHiP\,\cite{Anelli:2015pba} and MATHUSLA\,\cite{MATHUSLA:2022sze}, as well as LHCb-Run3 adopted from ref.\,\cite{Lanfranchi:2020crw}. We also include  constraints on the mediator from BBN \cite{Fradette:2017sdd} and SN1987a \cite{Krnjaic:2015mbs}. 

For complementary information, we highlight in fig.\,\ref{fig:scanID}, the subset of points within the projected reach of MATHUSLA (black), as well as those providing a good fit to the observed Galactic Centre excess (blue)\footnote{
The fit to the Galactic Centre excess (GCE), as well as the limits from indirect detection channels, are subject to astrophysical uncertainties. We therefore do not make strong claims regarding the exclusion of the model or a definitive explanation of the GCE. Instead, these results are to be viewed as indicative of the phenomenological potential of the scenario, showing that it can accommodate regions compatible with the GCE. With our goal being to complement this broader discussion with a quantitative analysis of the impact of non-thermal effects on the phenomenology. More robust future observations will clarify whether this parameter space warrants a dedicated analysis.
}
~\cite{Cholis:2021rpp}. Conversely in fig.\,\ref{fig:scanphi}, we indicate in red the points potentially accessible to indirect detection via CTA. Notably, there exist regions of parameter space that can be probed simultaneously by indirect detection and forward physics experiments, highlighting the complementarity of these searches.
We additionally highlight six benchmark points, indicated by \textcolor{magenta}{*}, chosen to exemplify different production modes and phenomenology. The corresponding parameter values are listed in section~\ref{sec:BMs} in tab.\,\ref{table}.

%%%%%%%%%%%%%%%%%%%%%%%%%%%%%%%%%%%%%%%%%%%%%%%%%%%%%%
\subsection{Reinterpretation of constraints from indirect detection and CMB}
\label{sec:ID}
For the indirect detection constraints, we reinterpret the published sensitivity projections from CTA~\cite{Acharyya:2020sbj} and the combined limits from a joint analysis of the gamma-ray flux from dwarf spheroidal galaxies (dSphs) performed by Fermi-LAT, MAGIC, HAWC, H.E.S.S. and VERITAS~\cite{Fermi-LAT:2025gei}. These limits were derived assuming direct annihilation of DM into two SM particles, producing photons with energy $E_\gamma = m_{\rm DM}$.

Since the relevant process in the present set-up is $SS \rightarrow \phi\phi \rightarrow \ldots \to 4\gamma$ , each annihilation produces twice as many photons with approximately half the energy per photon. Consequently, a constraint derived for a given DM mass in the direct annihilation case translates to a constraint at twice that mass in our scenario. While the modified photon multiplicity further introduces a rescaling of the cross-section limit by a factor of $1/2$.

The translation of limits on the observed flux $(\Phi)$ to that on the thermally averaged cross-section $\langle\sigma v\rangle$, using $\Phi \propto n_{\rm DM}^2 \langle\sigma v\rangle$, assumes that the DM number density corresponds to the observed relic abundance, $n_{\rm DM} \propto 0.12/m_{\rm DM}$. Rescaling the mass as above therefore requires a corresponding rescaling of the number density, giving an additional factor of $4$. Altogether, this gives
\begin{equation}
\label{eq:IDrescale}
\langle\sigma v\rangle^\textrm{constraint}_\textrm{sequential}(m)
= 4 \times \frac{1}{2} \times \langle\sigma v\rangle^\textrm{constraint}_\textrm{direct}(m/2).
\end{equation}

We also identify regions of parameter space giving the best fit to the observed Galactic Centre excess (GCE)  following a similar reinterpretation. The differential $\gamma$-ray flux is computed using the photon spectrum from PPPC4DMID~\cite{Cirelli:2010xx}, assuming an NFW density profile with local density $0.3~\mathrm{GeV}/\mathrm{cm}^3$ at $r_0 = 8.33$ kpc. The thermally averaged cross section $\langle \sigma v\rangle_{SS\rightarrow \phi\phi}$ is evaluated assuming a Maxwellian velocity distribution, and the flux is integrated over a $40^\circ \times 40^\circ$ region around the Galactic Centre. In accordance with eq.\,\eqref{eq:IDrescale}, the photon flux for mass $m_S$ is replaced by twice the flux evaluated at $m_S/2$.

We then perform a goodness-of-fit to the observed GCE spectrum~\cite{Cholis:2021rpp}, using
\begin{equation*}
    \chi^2_{GC}\equiv \sum\limits_{i=1}^{N_\text{data}} (th._i^2-obs._i^2) /{(N_\text{data}-1)}
\end{equation*}
where $obs.$ denotes the observed differential photon flux and $th.$ the corresponding theoretical prediction for a given parameter point.

While this recasting neglects potential subtleties associated with cascade kinematics, dedicated model independent studies indicate that the difference between direct and one-step cascade constraints is modest for the constraints that we consider (see fig.\,9 of ref.~\cite{Elor:2015bho}). Our procedure therefore provides a reliable estimate of the indirect detection bounds.

The constraints from CMB observations, in contrast, depend on the \textit{total} energy injection from DM annihilation and are largely insensitive to the details of the cascade, provided that the final states are SM particles (as is the case for our mediator $\phi$ eventually decaying into SM particles). The modified spectrum typically introduces corrections to the obtained limit on DM annihilation cross section of $\sim 10\%$ to $\sim 50\%$.

%%%%%%%%%%%%%%%%%%%%%%%%%%%%%%%%%%%%%%%%%%%%%%%%%%%%%
\section{Relic Density }
\label{sec:relic}
%%%%%%%%%%%%%%%%%%%%%%%%%%%%%%%%%%%%%%%%%%%%%%%%%%%%%%%

We now outline the different levels of description used to compute the relic abundance, corresponding to progressively more general treatments of the Boltzmann equation. These range from the standard number density evolution to a full phase-space treatment, and will form the basis for the comparisons performed in the following section.

The standard approach is based on a solution of the Boltzmann equation for the number densities of the dark sector particles, assuming that their phase-space distributions remain in kinetic equilibrium with the SM bath. We refer to this as the nBE treatment. Under this assumption, the shape of the distribution functions for all particles are fully specified by the SM bath temperature, and the dynamics can be described by a set of coupled equations for the yields $Y_\phi$ and $Y_S$:
\begin{eqnarray}
    \frac{\dd Y_\phi}{ \dd x}&=&\frac{s}{x\tilde{H}}\left(\langle\sigma v\rangle_{\phi\phi} (Y_{\phi,\eq}^2-Y_\phi^2) -\langle\sigma v\rangle_{\phi\phi\rightarrow SS} \biggl(Y_\phi^2-\frac{Y_{\phi,\eq}^2}{Y_{S,\eq}^2}Y_S^2\biggl)\right) \nonumber\\
    &&+~\frac{1}{x\tilde{H}s}\frac{m_\phi^2}{2\pi^2}\Gamma_{\phi\rightarrow\SM,\SM}T K_1\!\left(\frac{m_\phi}{T}\right)\left(1-\frac{Y_\phi}{Y_{\phi,\eq}}\right) \nonumber \label{eq:nBE1}\\
    &&+~\frac{1}{x\tilde{H}s}\frac{ m_h^2}{2\pi^2}\Gamma_{h \rightarrow \phi\phi}T K_1\!\left(\frac{m_h}{T}\right)\left(1-\frac{Y_\phi^2}{Y_{\phi,\eq}^2}\right)\theta_\textrm{EWPT}, \\
    \frac{\dd Y_S}{\dd x}&=&\frac{s}{x\tilde{H}}\left(\langle\sigma v\rangle_{SS}(Y_{S,\eq}^2-Y_S^2) -\langle\sigma v\rangle_{SS\rightarrow \phi\phi}\biggl(Y_S^2-\frac{Y_{S,\eq}^2}{Y_{\phi,\eq}^2}Y_\phi^2\biggl)\right) \nonumber \\
    &&+~\frac{1}{x \tilde{H}s}\frac{m_h^2}{2\pi^2}\Gamma_{h\rightarrow SS}TK_1\left(\frac{m_h}{T}\right)\left(1-\frac{Y_S^2}{Y_{S,\eq}^2}\right)\theta_\textrm{EWPT}, \label{eq:nBE2}
\end{eqnarray}
where $\theta_{\rm EWPT}\equiv \theta(x-x_{\rm EWPT}), \tilde{H}\equiv H/(1+\tilde{g})$ with $H$ the Hubble rate and $\tilde{g}\equiv \dd \ln g/\dd\ln x$. $g$ is the entropy degrees of freedom and $s$ is the entropy density. We define $x\equiv m_{\rm{DM}}/T$ with SM bath temperature $T=T_\SM$ and mass scale chosen as $m_{\rm{DM}}\equiv m_\phi$. The equilibrium yields are given by $Y_{i,\eq}=n_{i,{\rm eq}}/s= g_i m_i^2 T K_2(m_i/T)/(2\pi^2s)$. In both \eqref{eq:nBE1} and \eqref{eq:nBE2} the first line encodes effect of annihilation of $\phi\phi$ and $SS$ states to SM particles as well as of conversions $\phi\phi \leftrightarrow SS$, while the remaining lines come from decay processes.
Here, $\Gamma_{\phi\rightarrow \SM,\SM}$ denotes the total decay width of $\phi$ into SM particles. While this can be computed at tree level, such an approximation is not reliable at low mediator masses. We therefore use tabulated Higgs decay widths (including QCD corrections) with the HDECAY code~\cite{Djouadi:1997yw}, and determine the mediator lifetime via $\Gamma_\phi \approx \theta^2~\Gamma_{h\rightarrow \SM}(m_\phi)$. We also neglect inverse decays, as they are strongly suppressed.

In this integrated Boltzmann equation, all momentum dependence of the interaction rates is encoded in thermally averaged cross sections evaluated at the SM temperature, denoted $\langle\sigma v\rangle_{ii}$ for annihilations and $\langle\sigma v\rangle_{ii\rightarrow jj}$ for conversions, where $i\in \{\phi,S\}$ and $i\neq j$. While this approximation is well justified and gives accurate results when kinetic equilibrium is maintained, it is known to become inaccurate when the dark sector departs from equilibrium~\cite{Bringmann:2006mu,Binder:2017rgn,Binder:2021bmg}, as is the case in freeze-in production.\\

A more general treatment in such scenarios is obtained by tracking  the first \textit{two} moments of the distribution functions, leading to the coupled Boltzmann equations (cBE)~\cite{Bringmann:2006mu,Binder:2016pnr,Binder:2017rgn}. 
This extends the nBE approach, which retains only the zeroth moment, i.e.\ the number density. In the cBE framework, each species is therefore characterised by both a number density and an effective temperature, while assuming that the distribution function for each dark sector particle maintains an equilibrium form at its temperature.
The resulting system consists of a set of four coupled equations for $Y_i$ and $y_i$ where  $i \in \{\phi,S\}$. 
The equations for the yields have a structure similar to the nBE, but now depend explicitly on the dark sector temperatures $x_i = m_{\rm{DM}}/T_i = m_{\rm{DM}}^2/(y_i s^{2/3})$, taking the form:
\begin{eqnarray}
    \frac{\dd Y_i}{ \dd x}&=&\frac{s}{x\tilde{H}}\bigg(\langle\sigma v\rangle_{ii}(x) Y_{i,\eq}^2- \langle\sigma v\rangle_{ii}(x_i) Y_i^2 +\langle\sigma v\rangle_{jj\rightarrow ii}(x_j) Y^2_j - \langle\sigma v\rangle_{ii\rightarrow jj}(x_i) Y^2_i \bigg) \nonumber \\
    &&+~\frac{1}{x\tilde{H}s}\frac{m_i^2}{2\pi^2}\Gamma_{i\rightarrow \SM,\SM}\,T K_1\!\left(\frac{m_i}{T}\right)\left(1-\frac{Y_i}{Y_{i,\eq}}\right)  \nonumber\\
    &&+~\frac{1}{x\tilde{H}s}\frac{ m_h^2}{2\pi^2}\Gamma_{h \rightarrow i,i}\,T K_1\!\left(\frac{m_h}{T}\right)\left(1-\frac{Y_i^2}{Y_{i,\eq}^2}\right)\theta_\textrm{EWPT} .
\end{eqnarray}
The remaining two equations needed to close the system describe the temperature parameter evolution ($y_i\equiv \frac{m_i}{3 s^{2/3}} \left\langle \frac{p^2}{E} \right\rangle$)
\begin{eqnarray}
    \frac{\dd y_i}{ \dd x}&=&\frac{s}{x\tilde{H}}\bigg[
    \frac{Y_{i,\eq}^2}{Y_i}\Big(y_{i,\eq}\langle\sigma v\rangle_2^{ii}(x)-y_{i}\langle\sigma v\rangle^{ii}(x)\Bigl) +     y_i Y_i\Big(\langle\sigma v\rangle^{ii}(x_i)-\langle\sigma v\rangle_2^{ii}(x_i)\Big) \nonumber \\ 
    &&\qquad+ \frac{Y_{j}^2}{Y_i}\Big(y_{j}\langle\overline{\sigma v}\rangle_2(x_j)-y_{i}\langle\sigma v\rangle^{jj\rightarrow ii}(x_j)\Big) +
        y_i Y_i\Big(\langle\sigma v\rangle^{ii\rightarrow jj}(x_i)-\langle\sigma v\rangle_2^{i,i\rightarrow j,j}(x_i) 
   \Big) \bigg] \nonumber \\  
    &&+ \frac{1}{x\tilde{H}} \Gamma_{i\to \SM,\SM} \ y_i \bigl(I_0(x_i)-I_2(x_i)\bigl)
    \nonumber \\  
    &&+ \frac{1}{x\tilde{H}}\Gamma_{h \rightarrow ii}\frac{Y_h^{\eq}(x)}{Y_i}\Bigl(y_{i,\eq}(x)I_2^h(x)-y_i I_0^h(x) \Bigl)\theta_\textrm{EWPT}
     \nonumber \\    &&+\frac{y_i}{3T_i}\frac{H}{x \tilde H} \langle p^4/E^3\rangle_i \, ,
\end{eqnarray}
where the first two lines encode heat transfer between the SM bath, $\phi$ and $S$ populations due to annihilation and conversion processes, governed by the thermal averaged cross sections and its $p^2$
weighted analogue $\langle\sigma v\rangle_2$, while the remaining lines come from decay processes and 
expansion. The exact definitions of the functions used in the above equations can be found in Appendix \ref{sec:appa}.

The cBE approach leads to sufficiently accurate results for the relic abundance provided that either the distortions from thermal shape are mild, \textit{e.g.} due to efficient interactions with another particle or itself, or when the number changing processes are approximately momentum independent. If these conditions are not met, then a fully general description can be obtained by solving the Boltzmann equation for the phase-space distributions of the dark sector particles, without making any assumptions about their form.  
The full Boltzmann equation (fBE) for the evolution of the distribution functions $f_i$ in general have the form
\begin{equation}
\label{eq:beq1}
 E_{i}\left(\partial_t-H p\partial_p\right)f_{i}=
 \sum_{\textrm{proc.}}C_\textrm{proc.}[f_i] ,
\end{equation}
where $C_\textrm{proc.}[f_i]$ is the collision term for a given process, with the sum running over all processes modifying the distribution function $f_{i}$ for each particle $i$.
In order to study the evolution of the two dark sector particles, in particular to understand the effect of the out-of-equilibrium evolution of the mediator $\phi$ on the abundance of dark matter $S$, we must simultaneously track the evolution of the distribution functions for both.
This leads to two sets of coupled integro-differential equations:
\begin{eqnarray}
    E_\phi(\partial_t-Hp\partial p)f_\phi&=&C_\textrm{ann.}[f_\phi]+C_\textrm{h-decay}[f_\phi]+C_\textrm{conv.}[f_\phi,f_S]+C_{\phi-\textrm{decay}}[f_\phi], \label{eq:beq2}\\
    E_S(\partial_t-Hp\partial p)f_S&=&C_\textrm{ann.}[f_S]+C_\textrm{h-decay}[f_S]+C_\textrm{conv.}[f_S,f_\phi]. \label{eq:beq3}
\end{eqnarray}
Here, annihilations include the process $hh\leftrightarrow ii$ before EWPT and the Higgs mediated $\SM~\SM\leftrightarrow  ii$ after EWPT, for $ i \in \{\phi,S\}$. We neglect elastic scatterings of the dark sector particles with the SM bath, as these are suppressed for the feeble couplings relevant for freeze-in production. 

The coupled fBE are solved numerically after discretization on a suitably chosen momentum grid for each of the two dark sector particles. 
The implementation is carried out within the two-component version of the \texttt{DRAKE} code \cite{Binder:2021bmg} as developed in ref.\,\cite{Chatterjee:2025vdz}.
The collision terms, therefore retain the same structure as in ref.\,\cite{Chatterjee:2025vdz}, with the addition of the collision terms for decay processes,  expressions for which are given in Appendix~\ref{sec:appa}.\\

Several comments are in order regarding the implementation of the collision terms:

\begin{itemize}[label=--]
\item Throughout the analysis, equilibrium distributions are approximated by Maxwell Boltzmann statistics, justified by the highly dilute nature of the dark sector throughout the evolution.

\item For the conversion processes  $SS \leftrightarrow \phi\phi$, the amplitude receives contributions both from the four-point interaction ($\propto \lambda_{S\phi}^2$) and from an $s$-channel Higgs-mediated process ($\propto \lambda_{Sh}^2 \lambda_{\phi h}^2$), the latter arising from cubic interactions after electroweak symmetry breaking. However, this contribution is suppressed by higher powers of the small freeze-in couplings, $\lambda_{Sh}, \lambda_{\phi h} \ll 10^{-3}$, compared to the quartic coupling $\lambda_{S\phi} \sim \mathcal{O}(10^{-3} - 1)$ considered here.
We therefore approximate the conversion amplitude by retaining only the four-point contribution. This leads to a significant technical simplification of the collision term, in particular for the angular integrations, allowing the use of simplified expressions derived in~\cite{Aboubrahim:2023yag} and implemented in our previous work~\cite{Chatterjee:2025vdz}.

\item The collision terms for $\phi$ decays require the amplitudes for each decay channel, which is readily available at the tree-level. Tree level decays, however, are known to not be accurate for lower $\phi$ masses. While a fully consistent NLO implementation at the amplitude level is not available (being only available for inclusive decay widths).  We therefore adopt an implementation which retains the NLO corrected \textit{total} decay rate, while approximating the sum over the allowed decay channels.

To do this, we model the decay using a mock fermion $f$ with mass $1~\mathrm{MeV}$ and a small coupling $\lambda_y$ to $\phi$. The coupling is rescaled such that the resulting decay width matches $\Gamma_\phi$, obtained using tabulated Higgs decay widths.  This mock fermion,  with the rescaled coupling, is used to evaluate decay and inverse decay processes of $\phi$ in the fBE (eq.\,\eqref{eq:beq2}), while all $2\leftrightarrow2$ processes involving $f$ are set identically to zero. This procedure correctly reproduces the total decay rate, although it does not capture detailed kinematic features, which are only relevant near SM thresholds. We further neglect inverse decays as they are strongly suppressed and effectively kinematically disallowed once thermal masses of the SM particles are taken into account.

\item We also do not include self-scattering processes for the dark sector particles. Although large values of $\lambda_{S\phi}$ may suggest sizeable self-interactions, we find in the results section that in such cases (notably BM5 and BM6) the conversion processes alone are sufficient to thermalise the dark sector, leading to equalisation of the temperatures of $\phi$ and $S$. This is the same qualitative effect expected from strong self-scatterings.
\end{itemize}

%%%%%%%%%%%%%%%%%%%%%%%%%%%%%%%%%%%%%%%%%%%%%%%%%%%%%
\section{Results}
\label{sec:BMs}
%%%%%%%%%%%%%%%%%%%%%%%%%%%%%%%%%%%%%%%%%%%%%%%%%%%%%%

\begin{table*}[h]
\centering
\small
\setlength{\tabcolsep}{4pt} 
\resizebox{\textwidth}{!}{
\begin{tabular}{lcccccccccc}
\toprule
Name & $m_\phi$ & $m_S$ & $\theta$ & $\lambda_{h\phi}$ & $\lambda_{hS}$ & $\lambda_{S\phi}$ & $(\Omega h^2)_{\mathrm{nBE}}$ & $(\Omega h^2)_{\mathrm{cBE}}$ & $(\Omega h^2)_{\mathrm{fBE}}$ & Description \\
\midrule
BM1 & 20.3 & 30.4 & $1.09 \times 10^{-8}$ & $1.67 \times 10^{-13}$ & $4.80 \times 10^{-12}$ & $10^{-5}$ & 0.114 & 0.114 & 0.112 & direct FI \\
BM2 & 1.09 & 438. & $1.24 \times 10^{-5}$ & $3.20 \times 10^{-11}$ & $3.72 \times 10^{-13}$ & 0.155 & 0.122 & 0.0162 & 0.0420 & mixed + FPhys \\
BM3 & 4.66 & 586. & $4.15 \times 10^{-6}$ & $8.90 \times 10^{-11}$ & $4.32 \times 10^{-15}$ & 0.006 & 0.111 & 0.00065 & 0.0026 & seq. FI + FPhys \\
BM4 & 3.00 & 76.1 & $3.00 \times 10^{-5}$ & $8.00 \times 10^{-11}$ & $1.99 \times 10^{-13}$ & 0.030 & 0.124 & 0.0686 & 0.0804 & mixed + FPhys \\
\midrule
BM5 & 60.0 & 500. & $2.33 \times 10^{-9}$ & $6.14 \times 10^{-9}$ & $1.16 \times 10^{-12}$ & 0.28 & 0.121 & 0.0451 & -- & dark FO + CTA \\
BM6 & 60.0 & 80.1 & $1.87 \times 10^{-10}$ & $1.51 \times 10^{-9}$ & $2.20 \times 10^{-11}$ & 0.090 & 0.109 & 0.0468 & -- & dark FO + GCE \\
\bottomrule
\end{tabular}
}
\caption{The chosen benchmark points highlighting different phenomenological scenarios. The masses are given in GeV and the last column indicates dominant production modes: direct freeze-in (direct FI), sequential freeze-in (seq.~FI), re-annihilation followed by the freeze-out in the dark sector (dark FO), or mixed; as well as potential detection channels: in forward physics experiments (FPhys) and indirect detection in high energy $\gamma$-rays with CTA or possibly providing fit to the Galactic Centre excess (GCE).}
\label{table}
\end{table*}

%%%%%%%%%%%%%%%%%%%%%%%%%%%%%%%%%%%%%%%%%%%%%%%%%%%%%%%%%%%
\subsection{Benchmark points}
In this subsection, we analyse a set of representative benchmark points (BMs) selected from the scan, chosen for their phenomenological relevance, and spanning distinct dynamical regimes of DM production. The six BMs are indicated by $\textcolor{magenta}{*}$  in fig.\,\ref{fig:scanID} and fig.\,\ref{fig:scanphi}, with the corresponding parameter values listed in tab.\,\ref{table}.
The goal is to analyse the evolution of the dark sector particles
and to quantify the impact of their out-of-equilibrium dynamics on DM relic abundance.
To this end, we compare the evolution and final DM abundance obtained from the full phase-space Boltzmann  equation (fBE) and from the coupled number density and temperature equations (cBE), with that from the standard number density Boltzmann equation (nBE).

In the standard treatment, all dark sector particles are assumed to maintain kinetic equilibrium with the SM bath until their chemical decoupling processes are completed. Consequently, $\phi$ and $S$ share a common temperature with the SM bath, and their distribution functions are assumed to have equilibrium shapes at the SM temperature. 

\begin{figure}[t]
\centering
\includegraphics[scale=0.52]{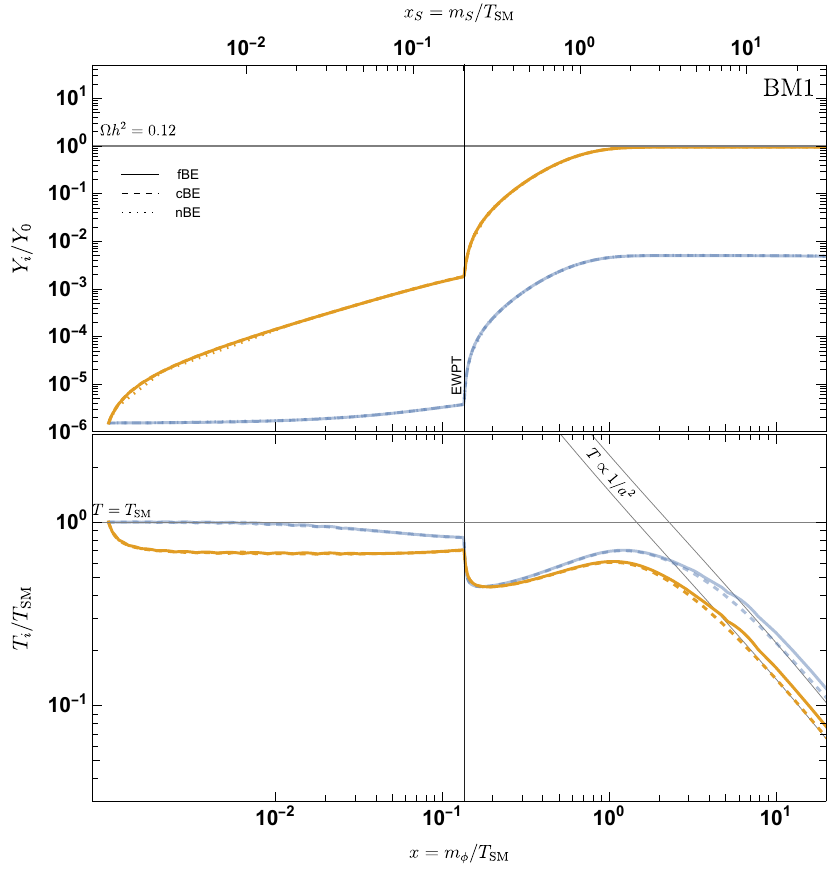}
\includegraphics[scale=0.52]{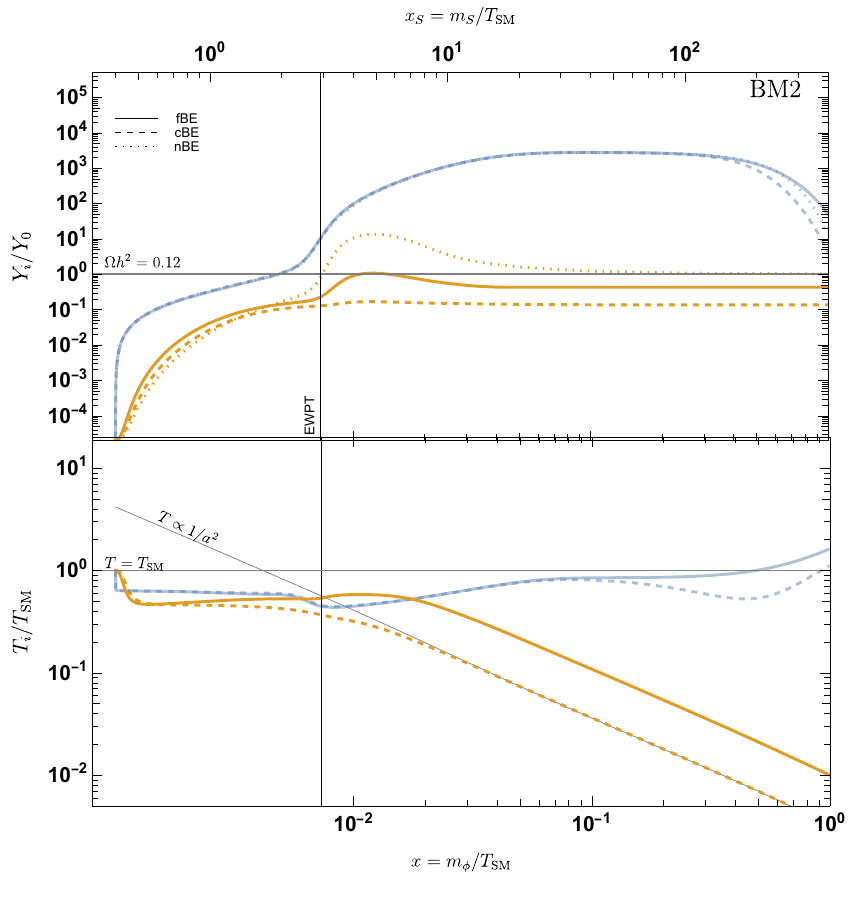}
\includegraphics[scale=0.52]{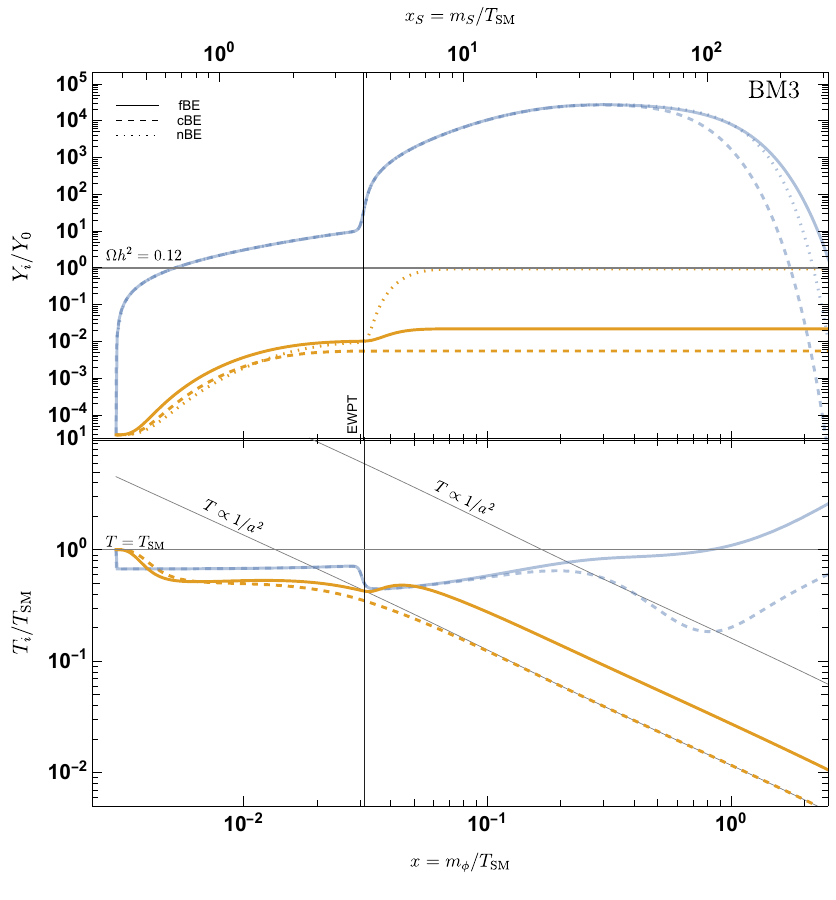}
\includegraphics[scale=0.504]{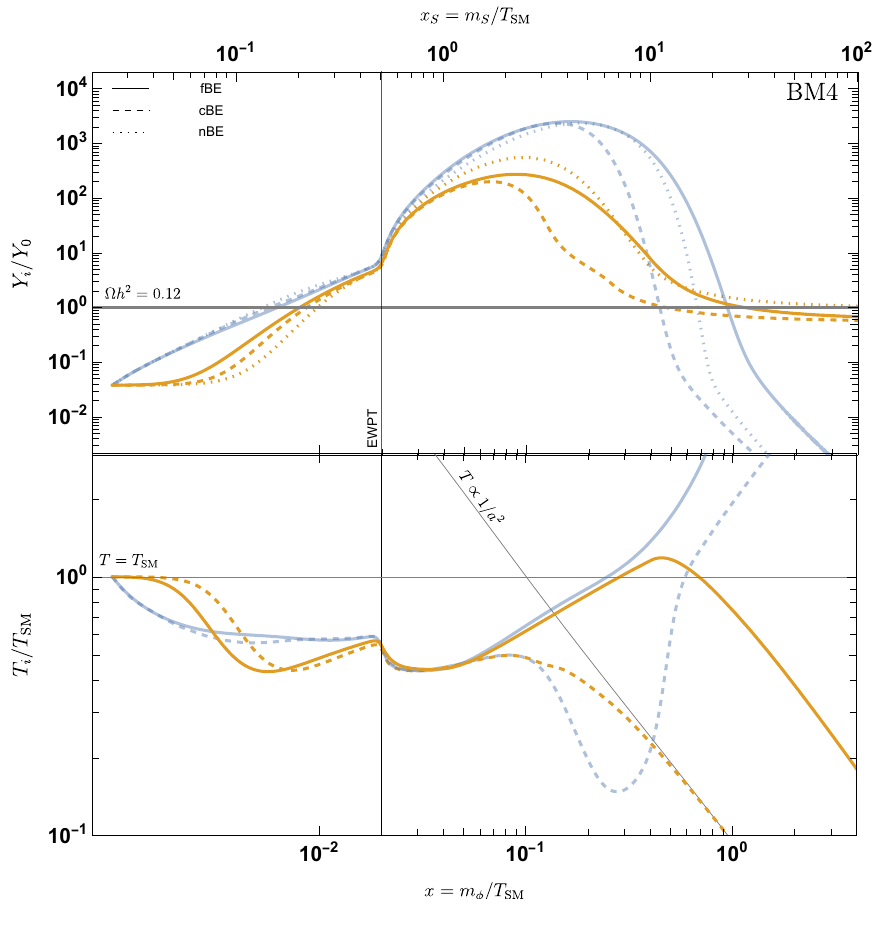}
\caption{
Impact of different Boltzmann equation treatments on the dark sector evolution for benchmarks BM1–BM4, characterised by dominantly freeze-in dynamics. \textbf{Upper panels:} Evolution of DM yields for $\phi$ (blue) and $S$ (orange) as obtained from the nBE (dotted), cBE (dashed), and fBE (solid) methods.
\textbf{Lower panels:} Temperature evolutions for $\phi$ (blue) and $S$ (orange) shown as the ratio of dark sector temperatures $(T_i)$ to the SM bath temperature $T_\SM$. This highlights the departure from the nBE assumption $(T_i=T_\SM)$ in the cBE (dashed) and fBE (solid) cases.}
\label{fig:BM14}
\end{figure}

A more general description is provided by the coupled Boltzmann equations (cBE),  which assumes that each dark sector particle maintains self-equilibrium during the chemical decoupling, but not necessarily kinetic equilibrium with the SM bath or with each other. In this case, $\phi$ and $S$ are allowed to have distinct temperatures, while their distribution functions are approximated by equilibrium forms at their respective temperatures.   Finally, the full Boltzmann equation makes no assumptions about the form of the distribution functions, allowing for a fully non-thermal phase-space evolution of both dark sector particles.

The cBE approach is widely employed in studies of extended dark sectors where kinetic equilibrium with the SM bath cannot be assumed.
A comparison between the cBE and the fBE treatment therefore isolates the effect of assuming thermal distributions at modified temperatures, as opposed to genuinely non-thermal phase-space distributions.

It is worth noting that the cBE solutions give an accurate approximation to the full phase-space result in the presence of large self-interactions \cite{Hryczuk:2022gay}. While for simplification, self-interactions are absent in the present setup, the cBE approximation is also expected to remain accurate when the conversion rates between the dark sector particles are large, establishing kinetic as well as chemical equilibrium within the dark sector. This is realised for the final two benchmark points discussed below.\\

As a first benchmark (BM1), we consider a \textit{direct} freeze-in scenario where the dark matter $S$ is produced directly from Higgs decays after EWPT, with no significant dynamical role of the mediator $\phi$ in its production. This is due to the relatively larger Higgs coupling to $S$, $\lambda_{Sh}\sim 30~\lambda_{\phi h}$, and a small conversion coupling.
The production thus corresponds to standard direct freeze-in, with no significant momentum dependence, and one sees in fig.\,\ref{fig:BM14} the nBE, cBE, and fBE results to be coinciding as expected. This also serves as a validation of our implementation of the different Boltzmann treatments.

\begin{figure}[t]
    \centering
    \includegraphics[width=0.69\linewidth]{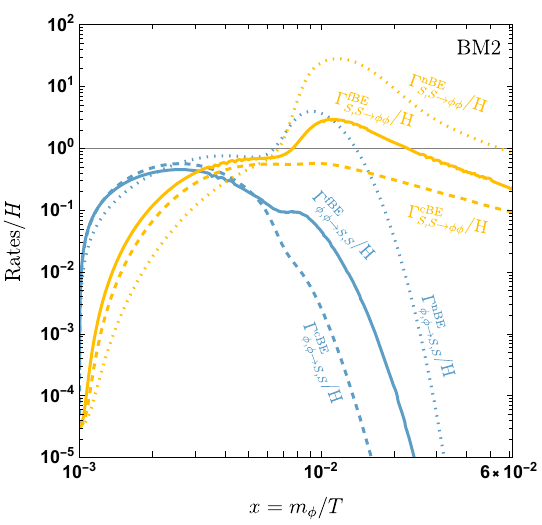}
    \caption{Conversion rates for BM2 from nBE, cBE and fBE solutions, extracted \textit{a posteriori}, as defined in eqs.\,\ref{eq:ratenBE}–\ref{eq:ratefBE}.}
    \label{fig:rplot}
\end{figure}

\begin{figure}[t]
\centering
\includegraphics[scale=0.785]{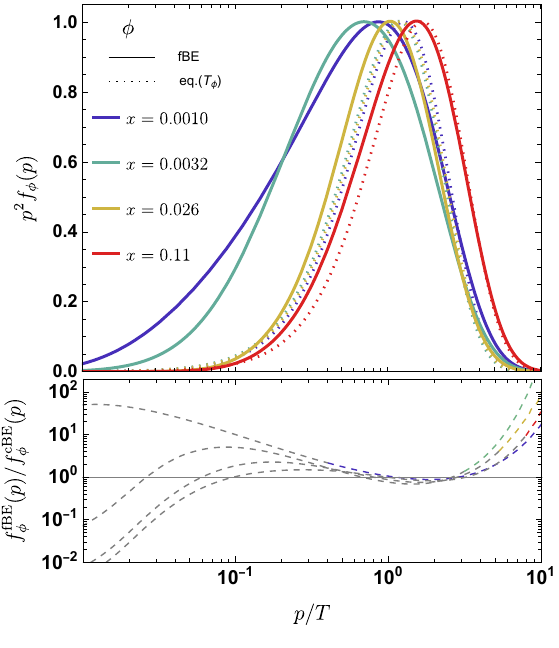}
\includegraphics[scale=0.70]{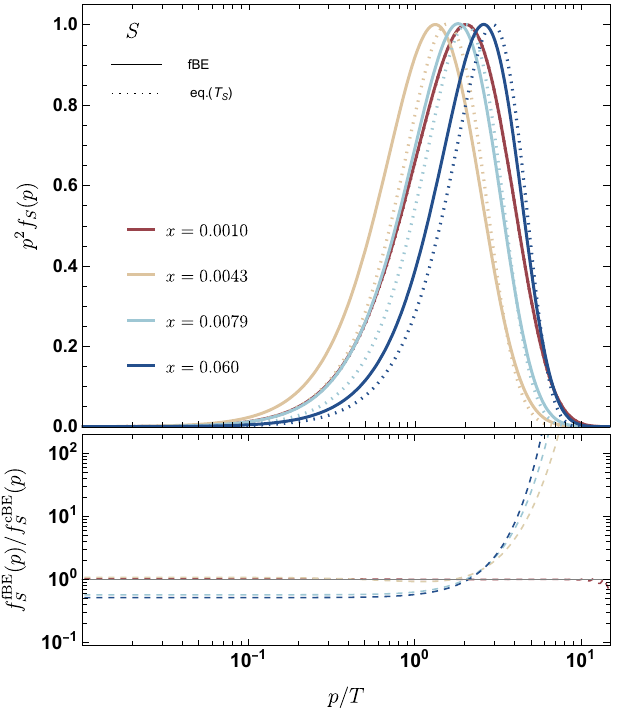}
\caption{In the upper panels, we show snapshots of the \textit{normalised} distribution functions of the dark sector particles at different times in the evolution for BM2. The solid lines correspond to the distributions obtained from the solution of the full Boltzmann equation (fBE). For comparison, the dotted lines indicate the corresponding equilibrium distributions evaluated at the effective temperatures extracted from the fBE solutions (shown by the solid lines in the lower panels of figs.\,\ref{fig:BM14} and \ref{fig:BM56}). The lower panels show the ratio between the two, highlighting the non-thermal \textit{shape} of the distribution functions captured only in the fBE treatment. To emphasise the parts of $f_\phi$ responsible for the observed deviation in the DM abundance, we shade in gray the low-momentum region that is kinematically forbidden from producing DM via conversions. The coloured regions therefore isolate the momentum range of $f_\phi$ that dominantly contributes to the deviation in the fBE abundance, as also discussed in the text.
}
\label{fig:bm2f}
\end{figure}

In BM2, we consider a  \textit{sequential} freeze-in scenario. We show the evolution of the yields and temperatures in the top and bottom panels for BM2 in fig.\,\ref{fig:BM14}, respectively. It proceeds as follows:
\begin{enumerate}[label=(\Roman*)]
    \item Initially, $\phi$ is produced from Higgs annihilations, with $S$  subsequently produced via $\phi\phi\rightarrow SS$ conversions. Accordingly, $Y_S$ trails behind $Y_\phi$. 
    \item After EWPT, Higgs decays become the dominant production channel for $\phi$. As the number density of $\phi$ increases, the conversion rate correspondingly grows, and the solutions for $Y_S$ obtained from nBE, cBE and fBE begin to deviate.

    This behaviour can be understood from the hierarchy of scales, $m_\phi< T_{\rm EWPT}<m_S$ with $m_S/m_\phi\sim \mathcal{O}(100)$, which implies that the conversion process $\phi\phi\rightarrow SS$ proceeds only through the high-momentum tail of the $\phi$ distribution, shortly after EWPT. In the nBE approach, $\phi$ is assumed to share the SM bath temperature, leading to an overestimation of the conversion rate. In contrast to the cBE treatment, which allows for a distinct temperature for $\phi$ and leads to solutions with $T_\phi<T_\SM$,  resulting in a suppressed conversion rate. In fig.\,\ref{fig:rplot}, we show the conversion rates, extracted \textit{a posteriori} from the solutions of the Boltzmann equations in all three treatments, defined as
    \begin{eqnarray}
        \Gamma^{\rm nBE}_{ii\rightarrow jj}&\equiv& n_i^{\rm nBE}\langle\sigma v\rangle_{ii\rightarrow jj}(x), \label{eq:ratenBE} \\
        \Gamma^{\rm cBE}_{ii\rightarrow jj}&\equiv& n_i^{\rm cBE}\langle\sigma v\rangle_{ii\rightarrow jj}(x_i^{\rm cBE}), \label{eq:ratecBE} \\
        \Gamma^{\rm fBE}_{ii\rightarrow jj}&\equiv& n_i^{\rm fBE}\langle\sigma v\rangle_{ii\rightarrow jj}^{f_i}(x), \label{eq:ratefBE}
    \end{eqnarray}
    with $x_i^{\rm cBE}\equiv m_{\rm DM}/T_i$, $i\neq j\in\{\phi,S\}$. And $\langle\sigma v\rangle_{ii\rightarrow jj}^{f_i}$ is the equivalent of the thermally averaged cross section found from the fBE solution, defined in Appendix \ref{sec:appa}.
    
    The fBE solution, which tracks the full momentum distribution of $\phi$ (and $S$) is found to yield a larger conversion rate for $\phi\phi\rightarrow SS$ than in the cBE case. This is despite the fact that the temperature $T_\phi$ inferred from fBE is found to overlap with that from cBE, shortly after EWPT. The difference instead arises from the non-thermal shape of the fBE distribution, which exhibits an enhanced high-momentum tail, as shown in fig.\,\ref{fig:bm2f}. And leads to a significant difference between the final abundances obtained from the cBE and fBE approaches, since the conversions can effectively \textit{only} proceed through this high momentum tail at this stage.

    \item Subsequently, around $x\sim0.01$, one observes a re-annihilation feature in the evolution of $S$ for the nBE and fBE solutions. This is due to the conversion rate in the inverse direction becoming large, $\Gamma_{SS \rightarrow \phi\phi}/H \gtrsim 1$. In contrast, in cBE one finds $\Gamma_{SS \rightarrow \phi\phi}/H \lesssim 1$, and the evolution of $S$ remains in the freeze-in regime. 

    \item At later times, $\phi$ decays deplete its abundance, leaving $S$ to constitute the entirety of the DM abundance.
\end{enumerate}

The impact of the non thermal distribution of the mediator particle $\phi$ on the DM abundance is two-fold.  
First, the kinematically suppressed conversion process selectively probes the high-momentum tail of the $\phi$ distribution. In the nBE treatment, this tail is effectively replenished by the assumption of kinetic equilibrium, leading to an overestimation of the conversion rate. In contrast, the fBE solution captures the depletion of this tail. Second, the temperature of $\phi$ differs from the SM bath temperature, as captured in the cBE and fBE approaches, further suppressing the conversion rate. The impacts, however, are not easily disentangled and the effect is cumulative.

Turning to the bottom panel of BM2, we consider the temperature evolution. At early times (prior to the EWPT), both $\phi$ and $S$ have temperatures of the same order as the SM bath temperature $T_\SM$. This reflects the fact that $2 \to 2$ processes produce particles with typical energies $\sim T$, and both species redshift similarly while relativistic ($T_{\SM} \gg m_\phi, m_S$).
Note that in fact $T_\phi$ and $T_S$ are smaller than $T_\SM$, which is due to the higher probability of interaction of the low energy part of the distribution in relativistic regimes $(\sigma_{2\rightarrow2}\propto 1/s)$. 

After the EWPT, Higgs decays $h \to \phi\phi$ initially lead to a drop in $T_\phi$ since $\phi$ particles are produced with energies $E_\phi \sim T_\SM/2$. While at later times, the  same Higgs decays produce $\phi$ particles with energies $E_\phi \sim m_h/2 \gg T_\SM$, resulting in a gradual increase in $T_\phi$ relative to $T_\SM$.
In contrast, $T_S$ decreases rapidly after the EWPT, 
with the decoupling of $S$ occurring when the inverse conversion rate satisfies $\Gamma_{SS \to \phi\phi}/H < 1$, and $T_S$ scales as $\sim1/a^2$. Consistent with the larger inverse conversion rates for fBE as compared to cBE, $S$ decoupling occurs relatively later in fBE.
Finally, the rise in $T_\phi$ at late times in both cBE and fBE coincides with the onset of $\phi$ decays, and is driven by the longer lifetimes of more energetic particles.

In fig.\,\ref{fig:bm2f}, we show snapshots of the momentum distribution functions of the two dark sector particles at representative values of $x$, as obtained from the fBE solution for BM2. For comparison, we also show the corresponding equilibrium distributions evaluated at the effective temperatures extracted from the fBE solutions. This allows for a direct assessment of deviations from equilibrium that arise purely from distortions in the \textit{shape} of the distributions, rather than from differences in temperature. Thus isolating genuine non-equilibrium effects in the phase-space distribution. One clearly observes that the distribution of $\phi$ departs from an equilibrium form, most notably through an enhanced high-momentum tail. The distortion of the distribution of S, in comparison, is rather mild. These deviations are further highlighted by the ratio to the corresponding equilibrium distributions shown in the lower panels.
\\

The benchmarks BM2--4 are chosen to probe parameter regions with potential sensitivity at forward physics experiments (as seen in fig.\,\ref{fig:scanphi}). In this mass range, the mediator can be long-lived, since the only kinematically allowed decays are to lepton pairs and photons.
Furthermore, for $m_S>m_h/2$, direct production of DM from Higgs decays is kinematically forbidden, with DM instead produced sequentially via $\phi$.
The three benchmarks however differ in the couplings and the mass of $S$, leading to  differences in the evolution and thereby quantitative differences in the non-thermal effect on DM abundance.
Nevertheless, some broad qualitative features of the evolution remain similar across these benchmarks.

In BM3, we observe again a production of DM via sequential freeze-in. The larger value of $m_S$, here, enhances the kinematic suppression of the conversion process. As a result, the production of $S$ becomes more sensitive to the high-momentum tail of the $\phi$ distribution, leading to a larger deviation between the abundances obtained from the standard nBE solution and those from the cBE/fBE treatments. The relevant control parameter is  $\sim m_S/T_\phi$ at the epoch when conversions freeze in. Additionally, the  evolution is simpler in this case as there is no later stage of re-annihilations, because the relatively smaller conversion coupling leads to conversion rates that never become larger than the Hubble rate.

While in BM4, the larger conversion coupling together with the smaller $m_S$ leads to efficient conversions in both directions, resulting in a thermalisation of the two dark sector particles. This is reflected in the temperature evolution, where $\phi$ and $S$ approach a common temperature distinct from the SM bath. The evolution exhibits a mixed freeze-in/freeze-out behaviour prior to decoupling (see fig.\,2 of \cite{Du:2021jcj}). BM4 also has the smallest lifetime of $\phi$ leading to an earlier decay.

The temperature evolutions in BM3 and BM4 closely resemble that of BM2, differing mainly in the values of the scales $m_S$, $m_h$ and $\Gamma_\phi$. Once the production of $\phi$ and $S$ become inefficient, their temperatures redshift as $1/a^2$. These regimes are marked by the end of Higgs decays ($T \simeq m_h/3$) and the slowdown of conversion processes, respectively. At later times, when $\Gamma_\phi/H \sim \mathcal{O}(1)$ and $\phi$ decays become efficient, the temperature of $\phi$ begins to rise again, as  as described for BM2.

\begin{figure}[h]
\centering
\includegraphics[scale=0.52]{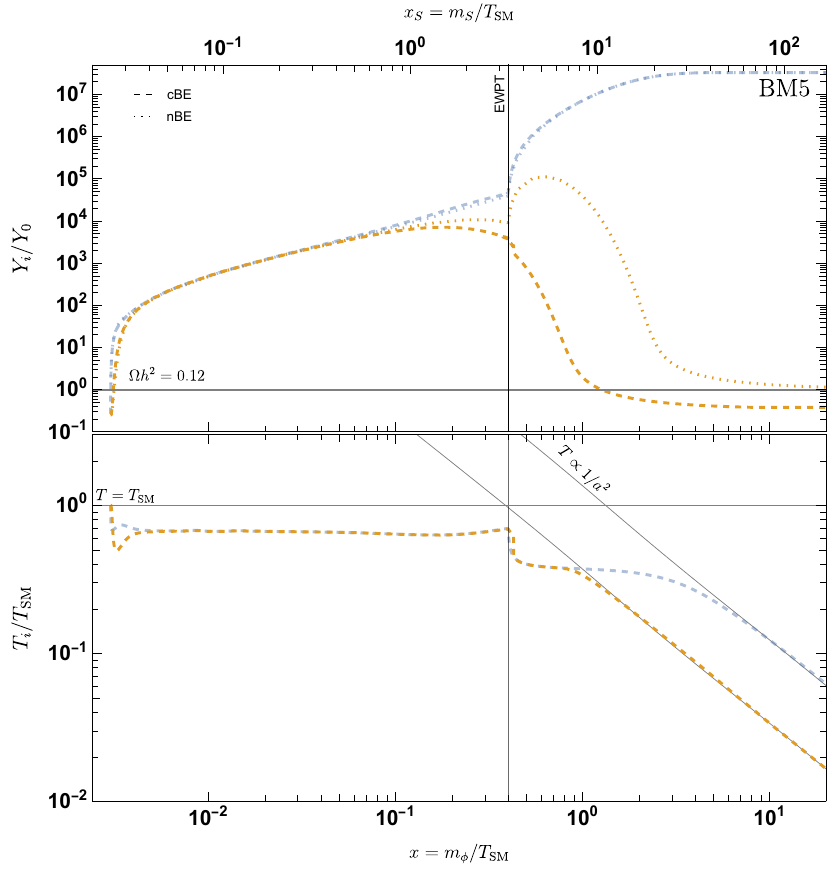}
\includegraphics[scale=0.52]{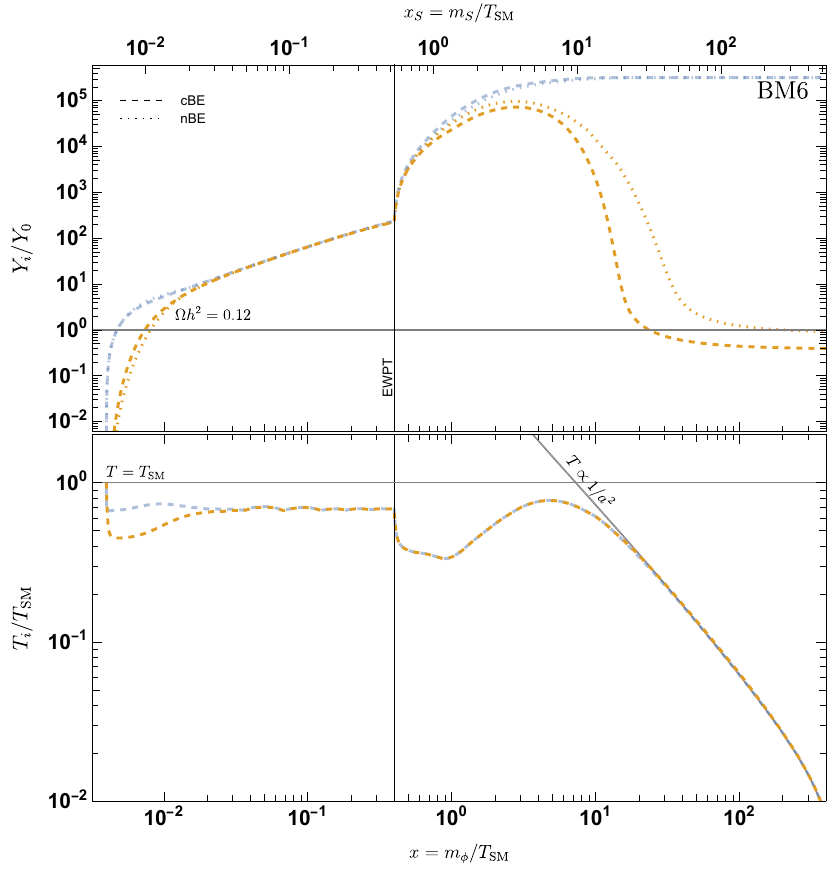}
\caption{Evolution of the yields $Y$ and temperatures for the benchmark points BM5 and BM6, exhibiting dark freeze-out.}
\label{fig:BM56}
\end{figure}

Finally, BM5 and BM6 are chosen with a focus on indirect detection, with prospective detection at CTA and as a potential explanation of the Galactic Centre Excess, respectively. 
Since the indirect detection goes through the channel $SS\rightarrow\phi\phi\rightarrow4\,\SM$, a significant flux requires a relatively larger value of conversion coupling $\lambda_{S\phi}$.
Therefore, the conversion rates are large enough for both these BMs, to rapidly thermalise the dark sector (even before significant production from Higgs decays occur). In this regime, the assumptions underlying the cBE method become effectively exact, with both $\phi$ and $S$ following equilibrium distributions at a common temperature $T_{\phi}=T_S \neq T_{\mathrm{SM}}$. Simultaneously, the numerical solution of fBE becomes challenging due to step size reduction when the large conversion rates enter an already steep equation.
However, since the distribution functions are driven close to equilibrium, the full phase-space treatment is not expected to provide additional physical insight, and we therefore restrict the discussion to nBE and cBE treatments for these benchmarks.

\begin{figure}[t]
    \centering
    \includegraphics[width=0.7\linewidth]{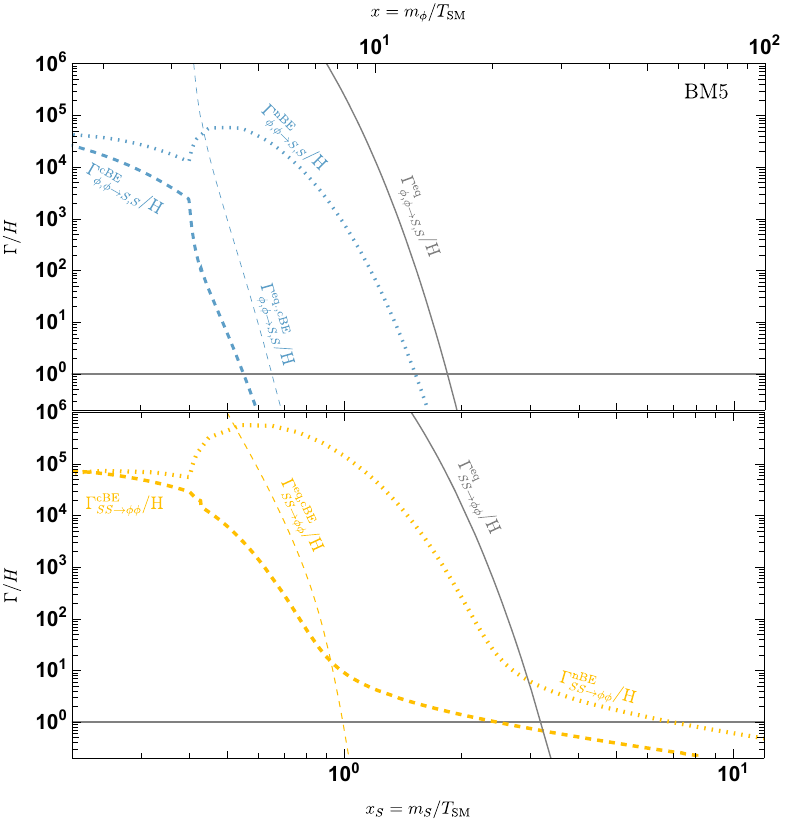}
    \caption{Conversion rates for BM5  from nBE and cBE solutions, extracted \textit{a posteriori}, as defined in eqs.\,\ref{eq:ratenBE}–\ref{eq:ratecBE}. Superimposed are the rates $\Gamma_{ii\rightarrow jj}^{\rm eq}(T_\SM)\equiv n_{i,{\rm eq}}\langle\sigma v\rangle_{ii\rightarrow jj}(T_\SM)$ and $\Gamma_{ii\rightarrow jj}^{\rm eq,~cBE}(T_\SM)\equiv n_{i,{\rm eq}}(T^{\rm cBE}_i)\langle\sigma v\rangle_{ii\rightarrow jj}(T^{\rm cBE}_i)$. For the process $\phi\phi\rightarrow SS$, the nBE and cBE rates flatten approximately where the corresponding equilibrium rates intersect them, analogous to the behaviour expected at freeze-out. Here, however, the freeze-out occurs in the conversion process within the dark sector, i.e.~exhibiting a dark freeze-out. }
    \label{fig:rplot5}
\end{figure}

For BM5, the DM abundance obtained in the cBE treatment is smaller than in nBE, primarily due to the lower mediator temperature 
$T_\phi<T_\SM$ when the dark sector temperatures are allowed to evolve independently. Combined with the strong sub-threshold suppression of the conversion process 
$\phi\phi\rightarrow SS$, this reduces the production of 
$S$, leading to the observed difference in  $Y_S$. This behaviour is also reflected in the rates shown in fig.\,\ref{fig:rplot5}. In the same figure, we also show the conversion rates computed assuming equilibrium distributions at the SM temperature and at the dark sector temperature obtained from the cBE solution, to be compared with the nBE and cBE rates, respectively. From the lower panel for $S$, one observes that the equilibrium rates cross the actual rates around the time when $\Gamma/H\sim \mathcal{O}(1)$, signalling a freeze-out like behaviour. This indicates that the production of $S$ in this case effectively proceeds via a dark freeze-out mechanism.

The evolution in BM6 is qualitatively similar. However, a good fit to the GCE signal favours a DM mass 
$m_S\sim\mathcal{O}(10\,\GeV)$, which is not significantly larger than the chosen  $m_\phi$. As a result, the sub-threshold suppression is reduced leading to the observed difference in $Y_S$ evolution, and the two dark sector particles undergo nearly simultaneous kinetic decoupling, leading to the observed differences in the temperature evolution.

%%%%%%%%%%%%%%%%%%%%%%%%%%%%%%%%%%%%%%%%%%%%%%%%%%%%%%%%%%%%
\subsection{Dependence on interaction strengths}
\label{sec:1D}

In the previous subsection, we analysed the evolution of the dark sector for a set of phenomenologically motivated benchmark points, relevant for detection at forward physics experiments and via indirect searches. Here, we extend the discussion to broader regions of parameter space to identify the generic features of the evolution of a two-component dark sector within the framework considered, and to demonstrate that the observed non-thermal effects are not tied to specific benchmark choices. As a result, we also analyse how these effects vary across the different production regimes.

Given the multidimensional nature of the parameter space, we adopt a simplified approach by performing one-dimensional variations in the couplings that dominantly control the number-changing processes and hence the relic abundance. These are the portal coupling $\lambda_{h\phi}$, which governs the injection of entropy from the SM bath into the dark sector, and the conversion coupling $\lambda_{S\phi}$, which controls the sequential production of DM.

\begin{figure}[h]
\centering
\includegraphics[scale=0.8]{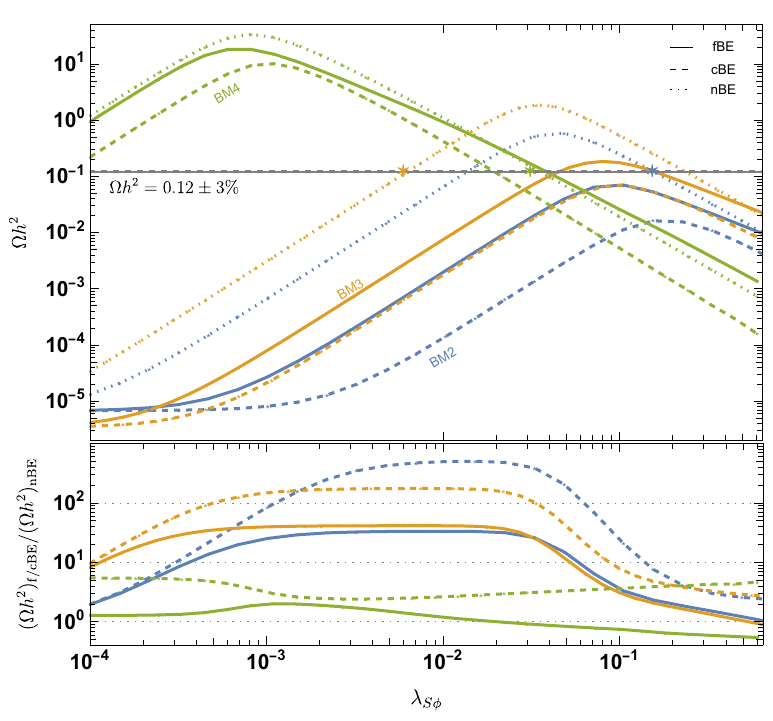}
\caption{Effect of varying the conversion strength on the resulting DM abundance, illustrated via three selected benchmark points. The BM points are indicated by $*$. In the lower panel, we show the ratios of the fBE and cBE solutions to the standard nBE solution, shown as solid and dashed lines, respectively.
}
\label{fig:1Da}
\end{figure}

\begin{figure}[h]
\centering
\includegraphics[scale=0.8]{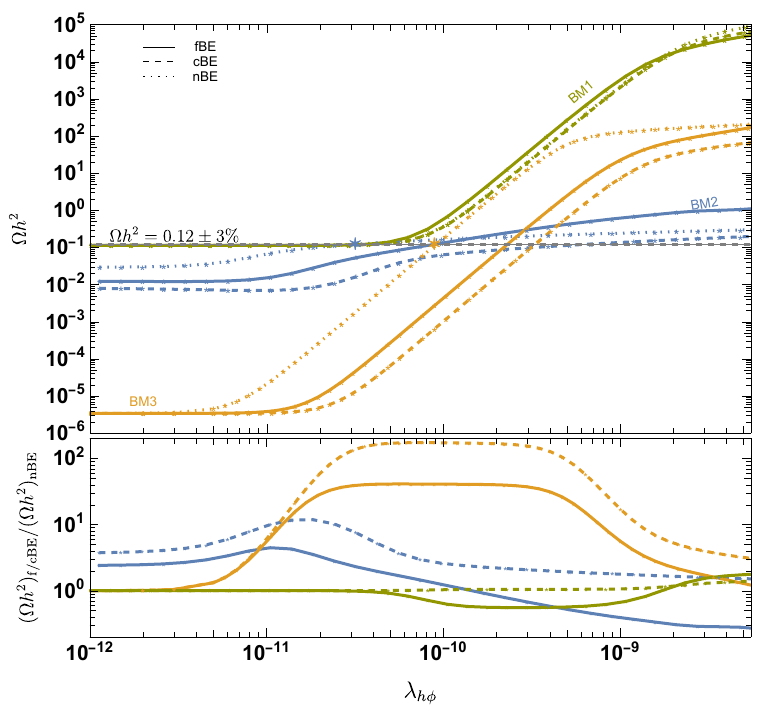}
\caption{Effect of varying the Higgs portal strength on the resulting DM abundance, illustrated via three selected benchmark points. The BM points are indicated by $*$ (with BM1 lying to the left of the plotted range). In the lower panel, we show the ratios of the fBE and cBE solutions to the standard nBE solution, shown as solid and dashed lines, respectively.
}
\label{fig:1Db}
\end{figure}

In fig.\,\ref{fig:1Da}, we vary the conversion coupling $\lambda_{S\phi}$, while holding all other parameters fixed. Across all three BMs, small $\lambda_{S\phi}$, the system remains in the sequential freeze-in regime, with non-thermal effects qualitatively understood as discussed in the previous section. Note that in the very small-$\lambda_{S\phi}$ limit, the conversions are strongly suppressed—particularly in the cBE and fBE treatments—rendering the DM abundance largely insensitive to the conversion strength, as reflected in the flattening of the curves. With increasing $\lambda_{S\phi}$, the conversion rate becomes comparable to the Hubble rate, leading to a transition from sequential freeze-in to a sub-threshold dark freeze-out regime, wherein the DM abundance is no longer set by sub-threshold freeze-in alone, and the impact of non-thermal effects on the DM abundance is reduced.  

In fig.\,\ref{fig:1Db}, we consider variations in the portal coupling $\lambda_{h\phi}$, with all other parameters fixed, focusing on BMs 1–3 indicated by the $*$ markers. Starting with BM1, at small $\lambda_{h\phi}$ the mediator abundance remains low and the system is in a \textit{direct} freeze-in regime, with nBE, cBE and fBE in agreement.
As $\lambda_{h\phi}$ increases, the mediator population becomes sufficiently large for the conversion process to become relevant.
In this regime, DM production is controlled by the high-momentum tail of the mediator distribution, and the non-thermal features of this distribution lead to the observed deviations the three treatments.

The behaviour of BM3 follows the same logic, only with the BM point falling at a different part of the 1D variation plot, as shown by the \textcolor{math2}{*}. 
In addition, the impact of the non-thermal mediator distribution is more pronounced due to the larger kinematic suppression of the conversion process $\phi\phi \rightarrow SS$, arising from the larger mass ratio $m_S/m_\phi$.
The nBE solution, which assumes a thermal distribution at the SM temperature, overestimates the relic abundance, while the difference between cBE and fBE reflects the role of the actual phase-space structure beyond a single effective temperature.
For BM2, on the other hand, the dependence on $\lambda_{h\phi}$ is relatively weak. This can be understood from its dark freeze-out–like evolution, where the DM abundance is primarily controlled by the conversion rate, and hence by the coupling $\lambda_{S\phi}$.\\

Together, the two plots show that the non-thermal effects captured by the fBE treatment exhibit a non-trivial dependence on the couplings of the model. This strengthens the case for dedicated studies of phase-space evolution in such scenarios, and further motivates the implementation of two-component freeze-in dynamics in numerical tools such as \texttt{DRAKE} or \texttt{BEST}~\cite{Yoon:2026rce}. The upcoming public release of \texttt{DRAKE} will include the framework for studying scenarios of the type discussed here.

%%%%%%%%%%%%%%%%%%%%%%%%%%%%%%%%%%%%%%%%%%%%%%%%%%%%%%%%%%%%
\section{Conclusions}
\label{sec:conclusions}
%%%%%%%%%%%%%%%%%%%%%%%%%%%%%%%%%%%%%%%%%%%%%%%%%%%%%%%%%%%%

The analysis presented in this work challenges a common assumption that particle production via the freeze-in mechanism is not affected by the details of their momentum distribution and thus that one can adopt thermal distribution functions with the temperature of the SM plasma. We performed a detailed case study of a minimal two-component dark sector model undergoing different realizations of the freeze-in production: direct, sequential and/or followed by a dark freeze-out. We find that only in the case of direct production from states in thermal equilibrium the assumption of thermal distribution functions with the temperature of the SM plasma is accurate, while whenever interactions between different dark sector states are relevant, departures from kinetic equilibrium can significantly modify not only the final momentum distribution, but also relic abundance, by even more than an order of magnitude. 

These results are worth keeping in mind when studying dark matter models with non-minimal freeze-in dynamics. We also provide a framework for how such effects can be taken into account. Furthermore, computational methods developed in this work constitute basis for implementation of two-component models freeze-in in the next public release of the \texttt{DRAKE} numerical code. 

Additionally, we show that the model studied in this paper, a two real scalar model coupled to the visible sector through a Higgs portal, is an interesting dark matter realization in its own right with potential for detection in the upcoming experimental searches, in particular forward physics experiments like MATHUSLA, SHiP and FASER and via potentially indirect detection in CTA. There also exists some complementarity between these detection methods, potentially strengthening dark matter interpretation of the observed signal. It is important to stress that in determining the predictions for the model's phenomenology, inclusion of departures from thermal equilibrium can be quite consequential for the abundances and subsequently the indirect detection signals, making them necessary to be taken into account.

%%%%%%%%%%%%%%%%%%%%%%%%%%%%%%%%%%%%%%%%%%%%%%%%%%%%%%%%%%%%
\acknowledgments
This work was supported by the National Science Centre (Poland) under the research Grant No. 2021/42/E/ST2/00009.

%%%%%%%%%%%%%%%%%%%%%%%%%%%%%%%%%%%%%%%%%%%%%%%%%%%%%%%%%%%%
\appendix
\section{Appendix}
\label{sec:appa}
The thermal averaged annihilation and conversion cross sections used in nBE and cBE are obtained via
\begin{equation}
\label{therm_av1}
 \left\langle \sigma v\right\rangle
 \equiv
\frac{g_i^2}{n_{i,{\rm eq}}^2}
\int \frac{d^3p}{(2\pi)^3} \frac{d^3\tilde p}{(2\pi)^3} 
 \sigma v_{ii\rightarrow \bar f f} 
f_{i,{\rm eq}}(\mathbf{p}) f_{i,{\rm eq}}(\tilde{\mathbf{p}})
\end{equation}
while the temperature-weighted thermal average for the $y$ evolution:
\begin{eqnarray}
 \left\langle \sigma v\right\rangle_{2}
 &\equiv& 
 \frac{g_i^2}{T_i n_{i,{\rm eq}}^2} \int  
 \frac{d^3p\, d^3\tilde p}{(2\pi)^6}
 \frac{p^{2}}{3E} 
 \sigma v f_{i,{\rm eq}}({\mathbf{p}}) f_{i,{\rm eq}}(\tilde{\mathbf{p}}) . 
\end{eqnarray}
In fBE treatment, one may also define similarly a thermally averaged cross section, but with the equilibrium quantities in eq.~\eqref{therm_av1} replaced by the values obtained from the fBE solution at each $x$-step:
\begin{equation}
\label{fBEtherm_av1}
 \left\langle \sigma v\right\rangle^f_i
 \equiv
\frac{g_i^2}{n_{i}^2}
\int \frac{d^3p}{(2\pi)^3} \frac{d^3\tilde p}{(2\pi)^3} 
 \sigma v f_{i}(\mathbf{p}) f_{i}(\tilde{\mathbf{p}}).
\end{equation}

Additionally we used a shorthand notation of
\begin{eqnarray}
        \langle\overline{\sigma v}\rangle_2(x_j) &\equiv& \frac{m_i}{m_j}\left( \frac{Y_{i,\eq}(x_j)}{Y_{j,\eq}(x_j)}\right)^2 \langle\sigma v\rangle_2^{ii\to jj}(x_j),
\end{eqnarray}
and the integrals needed for the decay contributions are given by
\begin{equation}
    \langle p^4/E^3\rangle_i\equiv \frac{g_i}{2\pi^2n_i^{\eq}(T_i)} \int dp \frac{p^6}{E_i^3} e^{-E_i/T_i},
\end{equation}
and terms contributing to the 2nd moment from decay processes are
\begin{eqnarray}
I_0(x)&\equiv&\frac{K_1(m_i/T_i)}{K_2( m_i/T_i)}, \\
I^h_0(x)&\equiv&\frac{K_1( m_h/T)}{K_2( m_h/T)}, \\
    I_2(x)&\equiv&\langle p^2/3E^2\rangle_i=\frac{g_i}{2\pi^2n_i^{\eq}(T_i)} \int dp \frac{p^4}{3E_i^2} e^{-E_i/T_i}, \\
    I^h_2(x)&\equiv&\frac{g_h}{2\pi^2n_h^{\eq}(T)}\frac{m_h}{T} \frac{1}{2\beta}\int dp \frac{p}{3E_h} e^{-E_h/T} \left[ q^2_{max\!}-q^2_{min}\!-m_i^2\log\frac{q^2_{max}+m_i^2}{q^2_{min}+m_i^2}\right]
\end{eqnarray}
with $\beta=\sqrt{1-4m_i^2/m_h^2}$ and $q_{max/min}=1/2|p\pm E_h\beta|$, $E_{i/h}=\sqrt{p^2+m^2_{i/h}}$.\\

The collision terms for decays in the fBE are:
\begin{eqnarray}
    C_{\rm h-decay}&\equiv& \frac{1}{2g_h} \frac{1}{x\tilde{H}}\int \dd\Pi_j \dd \Pi_h|\mathcal{M}|^2_{h\rightarrow ij} (2\pi)^4\delta^4\left(p_h-p_i-p_j\right)\left(f_h^{\rm eq}-f_if_j\right), \quad  \\
    C_{\phi-{\rm decay}}&\equiv&  \frac{-f_\phi}{16\pi |\vec{p}_\phi|g_\phi}\frac{1}{x\tilde{H}}\int_{E_2^-}^{E_2^+}\dd E_2|\mathcal{M}|^2_{\phi\rightarrow\SM\,\SM},
\end{eqnarray}
where $\Pi_i\equiv d^3p_i/(2\pi)^3(2E_i)$ and $E_2^\pm\equiv \biggl(E_\phi\pm |\vec{p}_\phi| \sqrt{1-4m_\SM^2/m_\phi^2}\biggl)\biggl/2$. We also ignore the inverse decays of $\phi$, as discussed in sec.~\ref{sec:relic}.

%%%%%%%%%%%%%%%%%%%%%%%%%%%%%%%%%%%%%%%%%%%%%%%%%%%%%%%%%%%%%%

\bibliography{biblio}{}
\bibliographystyle{JHEP}  

\end{document}